# Brillouin Klein space and half-turn space in three-dimensional acoustic crystals

Zhenxiao Zhu[1†], Linyun Yang[1†], Jien Wu[2†], Yan Meng[1], Xiang Xi[1], Bei Yan[1], Jingming Chen[1], Jiuyang Lu[2,3], Xueqin Huang[2], Weiyin Deng[2,3], Ce Shang[4], Perry Ping Shum[1], Yihao Yang[5], Hongsheng Chen[5], Gui-Geng Liu[6*], Zhengyou Liu[3,7*], Zhen Gao[1*]

[1]Department of Electronic and Electrical Engineering, Southern University of Science and Technology; Shenzhen 518055, China.

[2]School of Physics and Optoelectronics, South China University of Technology; Guangzhou, 510640, China

[3]Key Laboratory of Artificial Micro- and Nanostructures of Ministry of Education and School of Physics and Technology, Wuhan University; Wuhan 430072, China.

[4]King Abdullah University of Science and Technology (KAUST), Physical Science and Engineering Division (PSE), Thuwal 23955-6900, Saudi Arabia.

[5]Interdisciplinary Center for Quantum Information, State Key Laboratory of Modern Optical Instrumentation, ZJU-Hangzhou Global Science and Technology Innovation Center, College of Information Science and Electronic Engineering, ZJU-UIUC Institute, Zhejiang University; Hangzhou 310027, China.

[6]Division of Physics and Applied Physics, School of Physical and Mathematical Sciences, Nanyang Technological University; Singapore 637371, Singapore.

[7]Institute for Advanced Studies, Wuhan University, Wuhan 430072, China.

†These authors contributed equally to this work.

*Corresponding author. Email: gaoz@sustech.edu.cn (Z.G.); guigeng001@e.ntu.edu.sg (G.G.L.); zyliu@whu.edu.cn (Z.Y.L)

**The Bloch band theory[1] and Brillouin zone (BZ)[2] that characterize wave behaviors in periodic mediums are two cornerstones of contemporary physics ranging from condensed matter[3] to topological physics[4]. Recent theoretical breakthrough[5] revealed that, under the projective symmetry algebra enforced by artificial gauge fields, the usual two-dimensional (2D) BZ (orientable Brillouin two-torus) can be fundamentally modified to a non-orientable Brillouin Klein bottle with radically distinct topology and novel topological phases. However, the physical consequence of artificial gauge fields on the more general three-dimensional (3D) BZ (orientable Brillouin three-torus) was so far missing. Here, we report the first theoretical discovery and experimental observation of non-orientable Brillouin Klein space and**

orientable Brillouin half-turn space in a 3D acoustic crystal with artificial gauge fields. We experimentally identify peculiar 3D momentum-space non-symmorphic screw rotation and glide reflection symmetries in the measured band structures. Moreover, we demonstrate a novel 3D Klein bottle insulator featuring a nonzero $\mathbb{Z}_2$ topological invariant and self-collimated topological surface states at two opposite surfaces related by a nonlocal twist, radically distinct from all previous topological insulators. Our discovery not only fundamentally modifies the 3D Bloch band theory and 3D BZ, but also opens the door towards a wealth of previously overlooked momentum-space topologies and unexplored topological physics with gauge symmetry beyond the existing paradigms.

Since their introduction in 1929, the Bloch band theory[1] and Brillouin zone (BZ)[2] that characterize wave-like behaviors in periodic mediums have played a central role in condensed matter physics[3,4] and artificial crystals such as metamaterials[6-7], photonic crystals[8–15], acoustic crystals[16–25], electric circuits[26,27], mechanical networks[28–30], and cold-atom lattices[31]. Particularly, they are essential in the discovery and classification of topological phases of matter whose band topologies and topological invariants are usually defined on the two-dimensional (2D) BZ (Brillouin two-torus)[10–13,18–21] or three-dimensional (3D) BZ (Brillouin three-torus)[14,15,22–24]. For example, integrating the Berry curvature over the Brillouin two-torus surface yields the celebrated topological invariant "Chern number" which can be viewed as the number of monopoles of Berry flux inside a closed Brillouin two-torus surface[13,15,30,31]. Consequently, the BZ topology itself is crucial for the exploration of topological physics. Since the usual 2D BZ (3D BZ) is topologically an orientable Brillioun two-torus (three-torus), so far, the topological phases of matter are mainly characterized over the orientable Brillouin torus.

Meanwhile, gauge symmetry has modified physics in a fundamental way and greatly enriched the topological phases of matter[5,32–41]. For example, under $\mathbb{Z}_2$ gauge fields, i.e., hopping amplitudes with positive and negative signs, the algebraic structure of crystal symmetries will be projectively represented beyond the textbook space group theory and yields unprecedented novel topological band physics never witnessed under ordinary symmetries, such as Möbius-twisted topological phases[33–35], spinless mirror Chern insulators[36], topological phases switching[37–38], and high-order Stiefel-Whitney semimetals[39–40]. More remarkably, recent theoretical work[5] unveiled that the 2D Bloch band theory and 2D BZ can be fundamentally modified by the projective symmetry algebra enforced by artificial gauge fields. The fundamental domain of the 2D momentum space dramatically changed from the usual orientable Brillouin two-torus to a non-orientable Brillouin Klein bottle. The non-orientability of Brillouin Klein bottle results in peculiar 2D momentum-space non-symmorphic glide reflection symmetry and novel 2D Klein bottle insulators featuring two topological edge states related by a nonlocal twist. However, so far, the experimental observation of non-orientable Brillouin Klein bottle and Klein bottle insulators still remains elusive. More importantly, it is natural to ask what is the physical consequence of artificial gauge fields on the more general 3D Bloch band theory and 3D BZ. Does it generate any new topology that has never been witnessed in conventional 3D periodic physical systems?

Here, we report on the first experimental discovery of non-orientable Brillouin Klein space and orientable half-turn space in a 3D acoustic crystal with artificial gauge fields. Interestingly, we experimentally demonstrate that the previously discovered 2D Brillouin Klein bottle is only a special case of 3D Brillouin Klein space and half-turn space by cutting them appropriately. Moreover, we directly observe the unique non-symmorphic screw rotation and glide reflection symmetries in the 3D momentum space, which in turn thoroughly change the topology of the fundamental domain of 3D momentum space from the orientable Brillouin three-torus to non-orientable Brillouin Klein space or orientable Brillouin half-turn space. We further experimentally demonstrate a novel 3D Klein bottle insulator featuring a non-zero $\mathbb{Z}_2$ topological invariant and one pair of self-collimated topological surface states on parallel surfaces related by a nonlocal twist, radically distinct from all previous 3D topological insulators.

**Three fundamental three-manifolds**

We start with three different compact three-manifolds with no boundary. As shown in Fig. 1a, an orientable 3D torus, or three-torus, can be constructed from a cube by "gluing" its top and bottom, left and right, and front and back faces in a usual way so that the corresponding arrows marked on opposite faces match up. "Gluing" three pairs of opposite faces means that when a particle moving in the cube reaches a point on a face, it goes through it and appears immediately from the corresponding point on the opposite face, producing periodic boundary conditions along all three directions. Consequently, if there is a "T"-marked cube in the three-torus, an inhabitant of the three-torus would see infinite repeated images of the "T"-marked cube periodically arranged in a 3D cubic lattice, as illustrated in Fig. 1b.

In addition to the orientable three-torus, there are two other intriguing and fundamental three-manifolds in algebraic topology[42]: the non-orientable Klein space and the orientable half-turn space. As shown in Fig. 1c, the Klein space can be constructed from a cube by "gluing" its top and bottom, left and right faces in the usual way, while "gluing" its front and back faces with a side-to-side flip. If a particle moving in the cub reaches the back face, it'll return immediately from the front face mirror-reversed (reverse left and right side). Consequently, if there is a "K"-marked cube in the Klein space, an inhabitant of the Klein space looks up, down, left, or right, and he would see repeating images of the "K"-marked cube positioned just as they were in the three-torus. But when he looks back or straight ahead, the "K"-marked cube's nearest image appears to have undergone a side-to-side mirror reversal and interchange its left and right faces, as shown in Fig. 1d. A half-turn space is constructed in a similar way to the Klein space, but "gluing" the cube's front and back faces with a half-turn (180°) rotation, as shown in Fig. 1e. Therefore, if an inhabitant of the half-turn space looks back or straight ahead, the "Ht"-marked cube's nearest image appears to rotate 180º, as shown in Fig. 1f.

**Orientable Brillouin half-turn space**

It is known that a 3D BZ is topologically an orientable Brillouin three-torus. Remarkably, under artificial gauge fields, we discovered that the fundamental domain of the 3D momentum space can be topologically changed from the usual orientable Brillouin three-torus to a non-orientable Brillouin Klein space or orientable Brillouin half-turn space, which not only fundamentally modifies the 3D Bloch band theory and 3D BZ, but also results in unprecedented novel 3D topological states that are radically distinct from those defined over the usual Brillouin three-torus. To demonstrate the Brillouin Klein space and half-turn space, we first construct a 3D cubic lattice model with a unit cell consisting of four sites (beige spheres) coupled with positive (dark green cylinders) and negative (red cylinder) intralayer couplings, and the neighboring layers are coupled with positive interlayer chiral couplings (slanted dark green cylinders), as shown in Fig. 2a. The insets present the top view ($x$-$y$ plane) of the lattice model with each plaquette encloses a $\pi$ gauge flux and the front view ($x$-$z$ plane) of the lattice model with chiral interlayer couplings. Under the spatial mirror symmetry $M_x$ ($M_z$) in the $x$ ($z$) direction (dashed lines refer to the mirror planes), the distribution of gauge flux is invariant but the intralayer coupling sign (interlayer coupling configuration) is exchanged. Therefore, the projective representation of spatial reflection symmetries in the $x$ and $z$ directions is accompanied by a gauge transformation $G$, and the reflection symmetry is modified to $\mathbf{M}_{xz} = GM_{xz}$. While under the translation symmetry $L_y$ ($L_z$) with period $a$

(*h*) along the *y* (*z*) direction, both the gauge flux and the coupling sign keep unchanged. Consequently, the translation symmetry in the *y* (*z*) direction is not modified by the gauge transformation *G*, i.e., **L**$_y$ = *L*$_y$ (**L**$_z$ = *L*$_z$). Note that the gauge sign of each site will be exchanged when the gauge transformation *G* is translated *a* (*h*) in the *y* (*z*) direction, respectively, as indicated in the bottom panel of Fig. 2a. The algebraic relation will be projectively represented as

$$\mathbf{M}_{xz}\mathbf{L}_y\mathbf{M}_{xz}^{-1} = -\mathbf{L}_y, \text{ and } \mathbf{M}_{xz}\mathbf{L}_z\mathbf{M}_{xz}^{-1} = -\mathbf{L}_z^{-1}, \tag{1}$$

and the translation symmetry *L*$_y$ (*L*$_z$) is diagonalized as $\hat{L}_y = e^{ik_y a}$ ($\hat{L}_z = e^{ik_z h}$) along the *y* (*z*) direction in the 3D momentum space. Hence, the projective algebraic relation in equation (1) is equivalent to

$$\widehat{M}_{xz} e^{ik_y a} \widehat{M}_{xz}^{-1} = -e^{ik_y a} = e^{i(k_y + G_y/2)a}, \tag{2}$$

and

$$\widehat{M}_{xz} e^{ik_z h} \widehat{M}_{xz}^{-1} = -e^{-ik_z h} = e^{i(-k_z + G_z/2)h}, \tag{3}$$

where *G*$_y$ (*G*$_z$) is the length of the reciprocal lattice vector ***G***$_y$ (***G***$_z$). From equations (2) and (3), remarkably, we make a key discovery that $\widehat{M}_{xz}$ must contain a half translation in the reciprocal lattice along *k*$_y$ and *k*$_z$. The operator $\widehat{M}_{xz}$ is represented as

$$\widehat{M}_{xz} = U \mathcal{L}_{G_z/2} \mathcal{L}_{G_y/2} \widehat{m}_z \widehat{m}_x, \tag{4}$$

where $U = \tau_0 \otimes \sigma_1$ is a unitary matrix, $\tau$ and $\sigma$ denote two sets of Pauli matrices. $\widehat{m}_x$ ($\widehat{m}_z$) is the operator that inverses *k*$_x$ (*k*$_z$), $\mathcal{L}_{G_y/2}$ ($\mathcal{L}_{G_z/2}$) represents the half translation ***G***$_y$/2 (***G***$_z$/2) of the reciprocal lattice. Consequently, **M**$_{xz}$ can be regarded as a momentum-space screw rotation along the *k*$_y$ direction followed by a translation (***G***$_y$ + ***G***$_z$)/2.

Now we start elucidating the physical consequence of this peculiar 3D momentum-space non-symmorphic screw rotation symmetry enforced by the projective symmetry algebra. Considering the Hamiltonian *H*(***k***) (see Methods and Extended Data Fig. 1), the constraint from **M**$_{xz}$ in equation

(4) is

$$UH(k_x, k_y, k_z)U^\dagger = H(-k_x, k_y + \pi, -k_z + \pi), \tag{5}$$

where we assume that both *a* and *h* are equal to 1. Then

$$H(-k_x, k_y + \pi, -k_z + \pi)U|\psi(\mathbf{k})\rangle = E(\mathbf{k})U|\psi(\mathbf{k})\rangle, \tag{6}$$

which indicates that $U|\psi(\mathbf{k})\rangle$ is an eigenstate of $H(-k_x, k_y + \pi, -k_z + \pi)$ whose energy $E(\mathbf{k})$ is the same as that of an eigenstate $|\psi(\mathbf{k})\rangle$ of $H(\mathbf{k})$. Hence, the eigenenergy at $(k_x, k_y, k_z)$ is equivalent to that at $(-k_x, k_y + \pi, -k_z + \pi)$, indicating that the 3D BZ can be partitioned into two parts by a fixed $k_y$ (pink planes in Fig. 2f, g) and only one of them (half of the 3D BZ) is independent which can be regarded as the fundamental domain of the 3D momentum space.

This unique feature of 3D momentum-space non-symmorphic screw rotation symmetry can be observed in the spectrum of the 3D lattice model at fixed $k_y = -0.5\pi$ (Fig. 2b) and $k_y = -0.5\pi + \pi$ (Fig. 2c). As shown in Fig. 2d, the pink dashed lines can be obtained by a 180° rotation of the purple solid lines (constant energy cut in Fig. 2b) around the $k_y$ axis, which are exactly the pink solid lines (constant energy cut in Fig. 2c) in Fig. 2e, demonstrating the non-symmorphic screw rotation symmetry of the Bloch band structure. Note that this result holds true for any $k_y$ and $k_y + \pi$ (see Extended Data Fig. 2). Hence, by "gluing" the front ($k_y$) and back faces ($k_y + \pi$) of the half 3D BZ with a 180° rotation around the $k_y$ axis and the other two pairs of opposite faces in the usual way, the fundamental domain of the 3D momentum space (half 3D BZ) is topologically an orientable Brillioun half-turn space (Fig. 1e, f). Interestingly, by appropriately cutting the orientable Brillouin half-turn space, the obtained fundamental domain of the cut 2D BZ (half 2D BZ represented by cyan planes in Fig. 2f, g) with oppositely oriented boundaries along $k_y$ direction and periodic boundaries along $k_x$ ($k_z$) direction in Fig. 2f (Fig. 2g) can be "glued" together with the topology of a non-orientable Klein bottle. Under the cut half 2D BZ, the constant energy cuts

(purple solid lines in Fig. 2h, i) exhibit momentum-space glide reflection symmetry. After a reflection of the constant energy cut over the lower half 2D BZ through the $k_y$ axis (vertical grey dashed lines in Fig. 2h, i) followed by a half translation $\mathcal{L}_{G_y/2}$, it exactly coincides with the constant energy cut over the upper half 2D BZ. For comparison, in Fig. 2h, i, the constant energy cut over the lower half 2D BZ is translated to the upper half 2D BZ and marked as pink dashed lines. Interestingly, more non-orientable Brillioun Klein bottles can be cut from the orientable Brillioun half-turn space if the two oppositely oriented boundaries (magenta lines) of the fundamental domain of the cut 2D BZ (cyan plane) in Fig. 2f (Fig. 2g) move along opposite $k_z$ ($k_x$) directions with the same wave vector (see Extended Data Fig. 3).

**Non-orientable Brillouin Klein space**

Note that the whole system is protected by time-reversal symmetry, hence the Hamiltonian satisfies $H^*(-\boldsymbol{k}) = H(\boldsymbol{k})$. By combing equations. (5) and (6), the eigenenergy at ($k_x$, $k_y$, $k_z$) can be written as

$$E(k_x, k_y, k_z) = E(k_x, -k_y - \pi, k_z - \pi)$$
$$= E(-k_x, k_y + \pi, -k_z + \pi)$$
$$= E(k_x, \pi-k_y, k_z + \pi) \quad (7)$$

This indicates that the 3D BZ can also be partitioned into two parts by a fixed $k_z$ (pink planes in Fig. 3e, f) and only one of them (half of the 3D BZ) is independent as the fundamental domain of the 3D momentum space. Significantly, we observed momentum-space non-symmorphic glide reflection symmetry in the spectrum of the 3D lattice model at fixed $k_z = -0.5\pi$ (Fig. 3a) and $k_z = -0.5\pi + \pi$ (Fig. 3b). This unique feature is clearly illustrated in Fig. 3c, the pink dashed lines can be obtained by the reflection of the purple solid lines (constant energy cut in Fig. 3a) through $k_y =$

$0.5\pi$ (horizontal grey dashed line), which are exactly the pink solid lines (constant energy cut in Fig. 3b) in Fig. 3d, unambiguously demonstrating the 3D momentum-space glide reflection symmetry of the Bloch band structure. Thus, by "gluing" the bottom ($k_z$ plane) and top ($k_z + \pi$ plane) faces of the half 3D BZ with $k_y$ axis flip and other two pairs of opposite faces with the usual way, the fundamental domain of the 3D momentum space (half 3D BZ) is topologically a Klein space (Fig. 1c, d). For any fixed $k_x$, as shown in Fig. 3e, f and Extended Data Fig. 4a, b, c, the fundamental domain of the cut 2D BZ (cyan planes) is topologically a Klein bottle and the constant energy cuts in Fig. 3g, h and Extended Data Fig. 4d, e, f, also exhibit 2D momentum-space glide reflection symmetry. Counterintuitively, we can realize both orientable Brillouin half-turn space and non-orientable Klein space in the same 3D lattice model by partitioning the original 3D BZ into two parts along different ($k_y$ or $k_z$) directions due to the presence of gauge fields.

**Observation of Brillioun half-turn space and Klein space**

We start experimentally demonstrating the Brillouin half-turn space and Klein space in a 3D acoustic crystal. Acoustic crystals have hitherto provided a versatile platform to study various topological phases under the framework of quantum-classical analogies[25]. More importantly, the $\mathbb{Z}_2$ artificial gauge fields can be easily implemented in acoustic crystals by constructing positive and negative couplings[34-35,38,40-41]. As shown in Fig. 4a, our experimental sample, which consists of $30 \times 30 \times 15$ acoustic resonators, was fabricated by 3D printing with photopolymer materials. Each unit cell consists of four cylindrical acoustic resonators (beige color) with radius of $r_0$ and heights of $h_1$ and $h_2$, as illustrated in Fig. 4b. Straight tubes with radii of $r_1$ to $r_5$ connecting four acoustic resonators serve as the positive (dark green color) or negative (red color) intralayer couplings, and the dark green curved tubes with radius of $r_6$ serve as the positive interlayer chiral couplings. The whole structure is hollow with air and surrounded by hard walls. We first insert a broadband point-like sound source in the middle of the sample to excite the bulk states and then use a microphone probe to measure the complex acoustic pressure distributions (including amplitude and phase) within the sample. By performing 3D spatial Fourier transform to the measured complex acoustic pressure distributions from real space to reciprocal space, we can obtain the projected bulk band structures in the $k_x$-$k_y$, $k_y$-$k_z$, and $k_x$-$k_z$ planes, as shown in Fig. 4d-i. We first present the measured (color maps) and simulated (green lines) iso-frequency contours at 6.46kHz with fixed $k_y = -0.5\pi/a$ (orange-red plane in Fig.4c) and $k_y = -0.5\pi/a + \pi/a$ (blue plane in Fig. 4c), respectively, as shown in Fig. 4d, e. Remarkably, after rotating one of the two iso-frequency contours with 180° around the $k_y$ axis (normal vector of the $k_x$-$k_z$ plane), it will coincide exactly with the other one. Moreover, this conclusion applies to any fixed $k_y$ ranging from $-0.5\pi/a$ to $0.5\pi/a$ (see Extended Data Fig. 2), indicating that the $k_y = 0.5\pi/a$ plane (vertical pink planes in Fig. 2f, g and blue plane in Fig. 4c) divides the original 3D BZ into two half 3D BZs which are closely related by a translation of wavevector $k_y = \pi/a$ and a rotation of 180° around the $k_y$ axis. Therefore, only one of the two half 3D BZ is independent and can be viewed as the fundamental domain of 3D momentum space whose topology is an orientable Brillouin half-turn space. Moreover, by cutting the Brillouin half-turn space appropriately, we can obtain non-orientable Brillouin Klein bottle over which the band structure exhibits 2D momentum-space glide reflection symmetry. As shown in Fig. 4f, with fixed $k_x = 0$ (yellow plane in Fig. 4c), the reflection of the measured iso-frequency contour over the cut lower half 2D BZ $k_z \in [-0.5\pi/h, 1.5\pi/h) \times k_y \in [-0.5\pi/a, 0.5\pi/a)$ through the $k_z = 0.5\pi/h$ axis (vertical grey dashed line) coincides with that over the cut upper half 2D BZ $k_z \in [-0.5\pi/h, 1.5\pi/h) \times k_y \in [0.5\pi/a, 1.5\pi/a)$ after a half translation $\mathcal{L}_{G_y/2}$, which is exactly the unique property of Brillouin Klein bottle.

Besides the orientable Brillouin half-turn space, we can also observe a non-orientable

Brillouin Klein space in the same 3D acoustic crystal. As shown in Fig. 4g, h, after a reflection through the $k_y = 0.5\pi/a$ (horizontal grey dashed line), the measured (color maps) and simulated (green lines) iso-frequency contour at fixed $k_z = -0.5\pi/h$ (purple plane in Fig. 4c) in Fig. 4g coincides exactly with that at $k_z = -0.5\pi/h + \pi/h$ (pink plane in Fig. 4c) in Fig. 4h. Therefore, these two faces of the cut half 3D BZ ($k_z = -0.5\pi/h$ and $k_z = -0.5\pi/h + \pi/h$ planes) are closely related by a translation of wavevector $k_z = \pi/h$ and a reflection around the $k_x$ axis, unambiguously verify that the original 3D BZ can also be partitioned into two parts by $k_z = 0.5\pi/h$ plane (pink planes in Fig. 3e, f and Fig. 4c) and the fundamental domain of the 3D momentum space (half 3D BZ) is topologically a non-orientable Brillouin Klein space consisting of layered Brillouin Klein bottle[42] with arbitrarily fixed $k_x$ (see Extended Data Fig. 4). Indeed, as shown in Fig. 4i, the measured (color maps) and simulated (green lines) iso-frequency contour of the cut 2D BZ with fixed $k_x = 0.5\pi/a$ (cyan plane in Fig. 4c) exhibits similar 2D momentum-space glide reflection symmetry. Note that a single non-orientable Brillouin Klein space can also be realized by simply stacking a 2D lattice model with vertical interlayer couplings under artificial gauge fields (see Methods and Extended Data Fig. 5).

**Topological surface states of 3D Klein bottle insulator**

Finally, we explore the novel topological surface states of the 3D acoustic crystal with artificial gauge fields. Figure. 5a shows the calculated topological surface states dispersions on surface 1 (red sheet) and surface 2 (blue sheet) (parallel to y-z plane), respectively. To further identify the special features of the topological surface states dispersions, we plot in Fig. 5b the frequency-dependent surface states dispersions along $k_y$ direction with fixed $k_z = 0.5\pi/h$ (vertical grey plane in Fig. 5a). In this case, the fundamental domain of the cut 2D BZ (cyan plane in Fig. 2f) can be viewed as a Brillouin Klein bottle, over which the band structure exhibits 2D momentum-space glide reflection symmetry and leads to a nonlocal twist relation between the two topological surface states. It can be seen that, by translating $0.5\pi/a$ along the $k_y$ direction, the projected surface state dispersion of surface 1 (red line) coincides with that of surface 2 (blue line). This is because the non-symmorphic character of $\mathbf{M}_x$ in the 2D momentum space guarantees that only the projected surface state dispersions over $k_y \in [-\pi/2a, \pi/2a)$ are independent, while those over $k_y \in [\pi/2a, 3\pi/2a)$ can be deduced from the operation of $\mathbf{M}_x$. Specially, $\mathbf{M}_x$ nonlocally maps the topological surface state dispersion on one surface over $k_y \in [-\pi/2a, \pi/2a)$ to that on the other surface over $k_y \in [\pi/2a, 3\pi/2a)$. For the tight-binding model in Fig. 2a, a topological invariant defined on the cut half 2D BZ with $k_y \in [-\pi/2, \pi/2)$ and $k_x \in [-\pi, \pi)$ (cyan plane in Fig. 2f) can be given by

$$v = \frac{1}{2\pi}[\gamma(-\pi/2) + \gamma(\pi/2)] \mod 2. \qquad (8)$$

where $\gamma(-\pi/2)$ and $\gamma(\pi/2)$ are the Berry phases on the $k_y = -\pi/2$ and $k_y = \pi/2$ paths in the cut half 2D BZ, respectively. Considering the specific parameters for our model (see Methods and Extended Data Fig. 1), we obtain a nontrivial topological invariant $v = 1$, as shown in Fig. 5c. Moreover, this result holds true for all Brillouin Klein bottles in Extended Data Fig. 3a that are cut from the Brillouin half-turn space. Therefore, the 3D acoustic crystal can be termed as a 3D Klein bottle insulator with nontrivial $\mathbb{Z}_2$ topological invariant defined over the Brillouin half-turn space and supports novel topological surface states. To measure the acoustic field distributions and dispersions of the topological surface states, we place a point acoustic source (green star) at the center of surface 1 and surface 2 to excite the topological surface states and use a microphone probe to image their acoustic field distributions, as shown in Fig. 5d, both surfaces support self-collimated topological surface states. By performing Fourier transform to the measured acoustic field distributions, we obtain the measured topological surface states dispersions (color maps) of the two surfaces, as shown in Fig. 5e, f, matching well with the simulation results (green lines) and exhibiting 2D momentum-space glide reflection symmetry and nonlocal twisted correlation. To further explore the peculiar nonlocal twisted relation between the two surface states, we plot their simulated (Fig. 5g) and measured (Fig. 5h, i) iso-frequency contours. As shown in Fig. 5g, the reflection of the iso-frequency contours of surface 2 (blue dashed lines) through the axis of $k_z = 0.5\pi/h$ (horizontal grey dashed line) almost coincide with that of surface 1 (red solid lines) after a half translation $\mathcal{L}_{G_y/2}$, in which the slight mismatch results from the fact that the 3D tight-binding model is not fully mapped to the realistic 3D acoustic crystal (see Extended Data Fig. 6 for the tight-binding model results). More interestingly, topological surface states of the 3D Klein bottle insulator only exist on rotation-symmetry-breaking surfaces, rather than on rotation-symmetry-preserving surfaces as for conventional topological crystalline insulators. Specifically, topological surface states only exist on the two surfaces that are perpendicular to the *x*-axis (*y-z* plane), whereas the other four surfaces parallel to the *x*-axis (*x-z* or *x-y* plane) are gapped without any topological surface states because of their zero topological invariants (see Extended Data Fig. 7). Note that these unique characteristics of topological surface states in orientable Brillouin half-turn space also apply to the topological surface states in non-orientable Brillouin Klein space realized in a 3D lattice model with vertical interlayer couplings under artificial gauge fields (see Methods and Extended Data Fig. 8, 9).

**Discussion**

In conclusion, we have theoretically discovered and experimentally observed a non-orientable Brillouin Klein space and an orientable Brillouin half-turn space in a 3D acoustic crystal with artificial gauge fields. The interplay between artificial gauge fields and symmetry thoroughly changes the algebraic structure of crystalline symmetries, giving rise to peculiar momentum-space non-symmorphic screw rotation and glide reflection symmetries and fundamentally modifying the 3D Bloch band theory and 3D BZ. Under $\mathbb{Z}_2$ artificial gauge fields, we observed that the measured band structures exhibit unique screw rotation or glide reflection symmetry in the momentum space, which can reduce the original orientable Brillouin three-torus to orientable Brillouin half-turn space or non-orientable Brillouin Klein space, and consequently changes the topological classification from the bottom level. Moreover, we experimentally demonstrate a novel 3D Klein bottle insulator featuring nonzero topological invariant and topological surface states at two opposite surfaces related by a nonlocal twist. Our discovery opens a new avenue to explore the novel topologies of momentum space and the unexplored gauge-symmetry-enriched topological physics that beyond the scope of topological quantum materials. We envision that other novel momentum-space topologies, such as quarter-turn space[42], hexagonal torus[42], and Roman surface[43] can also be realized in acoustic crystals with artificial gauge fields.

**References**

1. Bloch, F. Über die Quantenmechanik der Elektronen in Kristallgittern. *Z. Physik* **52**, 555–600 (1929).
2. Brillouin, L. Les électrons libres dans les métaux et le role des réflexions de Bragg. *J. Phys. Radium* **1,** 377–400 (1930).
3. Giustino, F. Electron-phonon interactions from first principles. *Rev. Mod. Phys.* **91**, 019901 (2019).
4. Bansil, A., Lin, H. & Das, T. Colloquium: Topological band theory. *Rev. Mod. Phys*. **88**, 021004 (2016).
5. Chen, Z. Y., Yang, S. A. & Y. X. Zhao, Brillouin Klein bottle from artificial gauge fields. *Nat. Commun*, **13**, 2215 (2022).
6. Ma, S. J., Yang, B. & Zhang, S. Topological photonics in metamaterials. Photon. Insights. **1**, R02 (2022).
7. Ni, X., Yves, S., Krasnok, A. & Alu, A. Topological metamaterials. Preprint at https://arxiv.org/abs/2211.10006 (2022).

8. Yablonovitch, E. Inhibited spontaneous emission in solid-state physics and electronics. *Phys. Rev. Lett.* **58**, 2059–2062 (1987).
9. John, S. Strong localization of photons in certain disordered dielectric superlattices. *Phys. Rev. Lett.* **58**, 2486–2489 (1987).
10. Haldane, F. D. M. Model for a quantum Hall effect without Landau levels: condensed-matter realization of the "parity anomaly". *Phys. Rev. Lett*. **61**, 2015–2018 (1988).
11. Wang, Z., Chong, Y., Joannopoulos, J. D. & Soljačić, M. Observation of unidirectional backscattering-immune topological electromagnetic states. *Nature* **461**, 772–775 (2009).
12. Lu, L., Joannopoulos, J. D. & Soljačić, M. Topological photonics. *Nat. Photon.* **8**, 821–829 (2014).
13. Ozawa, T. et al. Topological photonics. *Rev. Mod. Phys.* **91**, 015006 (2019).
14. Yang, Y. et al. Realization of a three-dimensional photonic topological insulator. *Nature* **565**, 622–626 (2019).
15. Liu, G. et al. Topological Chern vectors in three-dimensional photonic crystals. *Nature* **609**, 925–930 (2022).
16. Kushwaha, M., Halevi, P., Dobrzynsi, L. & Djafari-Rouhani, B. Acoustic band structure of periodic elastic composites. *Phys. Rev. Lett.* **71**, 2022–2025 (1993).
17. Yang, Z. J. et al. Topological Acoustics. *Phys. Rev. Lett.* **114**, 114301 (2015).
18. Khanikaev, A. B., Fleury, R., Mousavi, S. H. & Alù, A. Topologically robust sound propagation in an angular-momentum-biased graphene-like resonator lattice. *Nat. Commun.* **6**, 8260 (2015).
19. He, C. et al. Acoustic topological insulator and robust one-way sound transport. Nat. Phys. **12**, 1124–1129 (2016).
20. Ding, Y. et al. Experimental demonstration of acoustic Chern insulators. *Phys. Rev. Lett.* **122**, 014302 (2019).
21. Hu, B. et al. Non-Hermitian topological whispering gallery. *Nature* **597**, 655–659 (2021).
22. He, H. et al. Topological negative refraction of surface acoustic waves in a Weyl acoustic crystal. *Nature* **560**, 61–64 (2018).
23. He, C. et al. Acoustic analogues of three-dimensional topological insulators. *Nat. Commun.* **11**, 2318 (2020).
24. Lin, Z. K. et al. Topological Wannier cycles induced by sub-unit-cell artificial gauge flux in a sonic crystal. *Nat. Mater.* **21**, 430–437 (2022).
25. Xue, H., Yang, Y. & Zhang, B. Topological acoustics. *Nat. Rev. Mater.* **7**, 974–990 (2022).
26. Ningyuan, J., Owens, C., Sommer, A., Schuster, D. & Simon, J. Time-and site-resolved dynamics in a topological circuit. *Phys. Rev. X* **5**, 021031 (2015).
27. Imhof, S. et al. Topolectrical-circuit realization of topological corner modes. *Nat. Phys*. **14**, 925–929 (2018).

28. Wang, P., Lu, L. & Bertoldi, K. Topological phononic crystals with one-way elastic edge waves. *Phys. Rev. Lett.* **115**, 104302 (2015).

29. Huber, S. D. Topological mechanics. *Nat. Phys.* **12**, 621–623 (2016).

30. Ma, G., Xiao, M. & Chan, C. T. Topological phases in acoustic and mechanical systems. *Nat. Rev. Phys*. **1**, 281–294 (2019).

31. Cooper, N. R., Dalibard, J. & Spielman, I. B. Topological bands for ultracold atoms. *Rev. Mod. Phys.* **91**, 015005 (2019).

32. Shao, L. B., Liu, Q., Xiao, R., Yang, S. A. & Zhao, Y. X. Gauge-field extended $k \cdot p$ method and novel topological phases. *Phys. Rev. Lett.* **127**, 076401 (2021).

33. Zhao, Y. X., Huang, Y. X. & Yang, S. A. $Z_2$-projective translational symmetry protected topological phases. *Phys. Rev. B* **102**, 161117(R) (2020).

34. Xue, H. et al. Projectively enriched symmetry and topology in acoustic crystals. *Phys. Rev. Lett.* **128**, 116802 (2022).

35. Li, T. et al. Acoustic Möbius insulators from projective symmetry. *Phys. Rev. Lett.* **128**, 116803 (2022).

36. Shao, L., Chen, Z., Wang, K., Yang, S. A. & Zhao, Y. Spinless mirror Chern insulator from projective symmetry algebra. Preprint at https://arxiv.org/abs/2207.05654 (2022).

37. Zhao, Y. X., Chen, C., Sheng, X.-L. & Yang, S. A. Switching spinless and spinful topological phases with projective PT symmetry. *Phys. Rev. Lett.* **126**, 196402 (2021).

38. Meng, Y. et al. Spinful topological phases in acoustic crystals with projective PT symmetry. *Phys. Rev. Lett*. **130**, 206101 (2023).

39. Chen, Z. Y., Zhang, Z., Yang, S. A. & Zhao, Y. X. Classification of time-reversal-invariant crystals with gauge structures. *Nat. Commun.* **14**, 743 (2023).

40. Xue, H. et al. Stiefel-Whitney topological charges in a three-dimensional acoustic nodal-line crystal. Preprint at https://arxiv.org/abs/2304.06304 (2023).

41. Xiang, X. et al. Demonstration of acoustic high-order Stiefel-Whitney semimetal in bilayer graphene sonic crystals. Preprint at https://arxiv.org/abs/2304.12735 (2023).

42. Weeks, J. R. *The Shape of Space*. (CRC Press, 2020).

43. Liu, G. et al. Physical realization of topological Roman surface by spin-induced ferroelectric polarization in cubic lattice. *Nat. Commun.* **13**, 2373 (2022).

**Main figure legends**

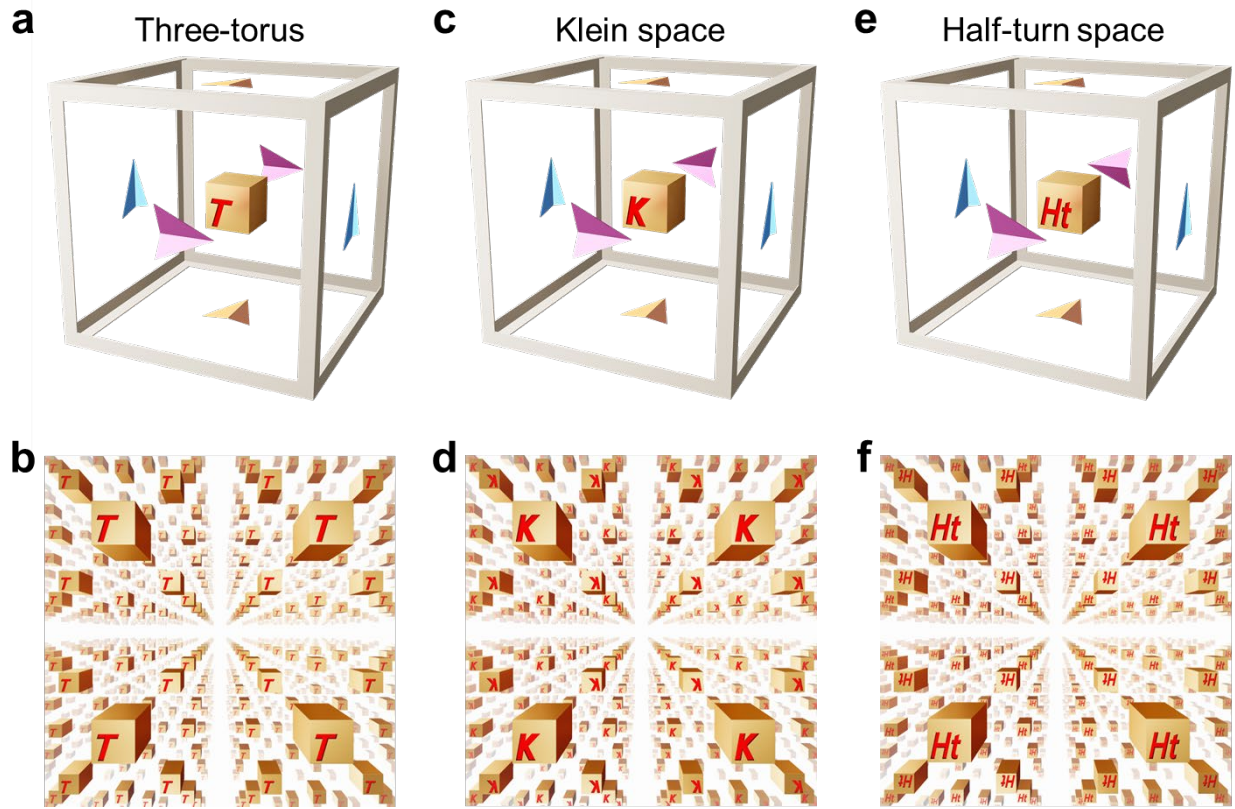

**Fig. 1 | Three fundamental three-manifolds. a**, An orientable three-torus is formed by gluing three pairs of opposite faces of a cube in the usual way. **b**, The view inside a three-torus containing a single "T"-marked cube. **c**, A non-orientable Klein space is formed by gluing the top to the bottom, the left to the right faces in the usual way, but the front to the back faces with a side-to-side flip. **d**, The view inside a Klein space containing a single "K"-marked cube. **e**, An orientable half-turn space is formed by gluing the top to the bottom, the left to the right faces in the usual way, but the front to the back faces with a 180º rotation. **f**, The view inside a half-turn space containing a single "Ht"-marked cube. The direction and the light and dark color of the purple arrows indicate the way of gluing the front to the back faces.

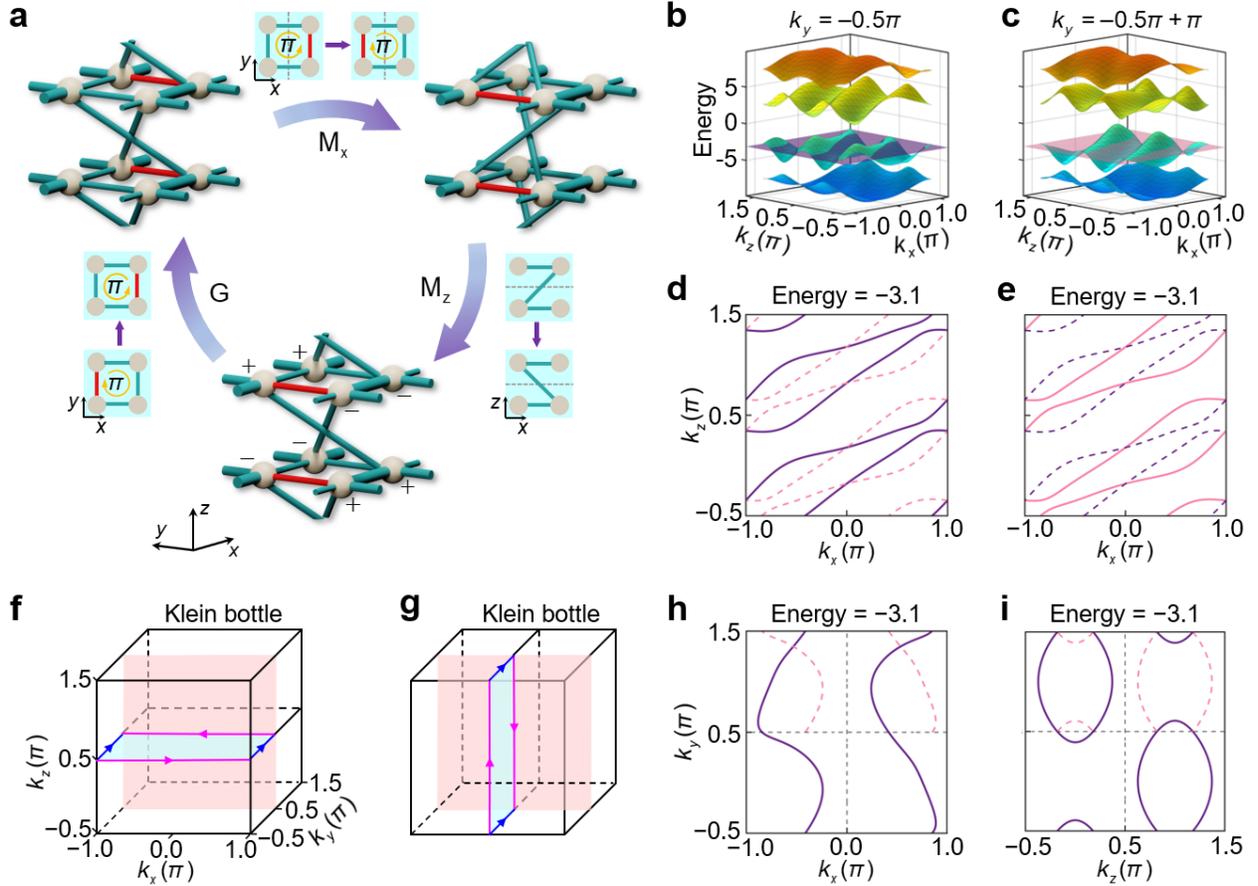

**Fig. 2 | Brillouin half-turn space with momentum-space screw rotation symmetry. a,** Schematic of the cubic lattice model consisting of two unit-cells in the $z$ direction. The tight-binding configuration is invariant under the spatial mirror operations $M_x$ and $M_z$ in the $x$ and $z$ directions and the gauge transformation G. The green (red) cylinders indicate positive (negative) couplings. **b, c,** Energy bands of the model with fixed $k_y = -0.5\pi$ and $k_y = -0.5\pi + \pi$. **d, e,** The constant energy cuts that correspond to the purple plane in **b** and the pink plane in **c**, respectively. In **d** and **e**, the pink (purple) dashed lines can be obtained by rotating the purple (pink) solid lines in **d** (**e**) with 180° around the $k_y$ axis, which are exactly the pink (purple) solid lines in **e** (**d**), demonstrating the momentum-space screw rotation symmetry of the band structures in the Brillouin half-turn space. **f, g,** Brillouin Klein bottles (cyan planes) can be achieved by appropriately cutting the Brillouin half-turn space. The marked arrow directions indicate how the boundaries with the same color will be "glued" together. **h, i,** The constant energy cuts correspond to the cut 2D BZ planes in **f, g,** respectively, which exhibit momentum-space glid reflection symmetry. For comparison, the constant energy cuts within the lower half 2D BZ is translated to the upper half 2D BZ and marked as pink dashed lines.

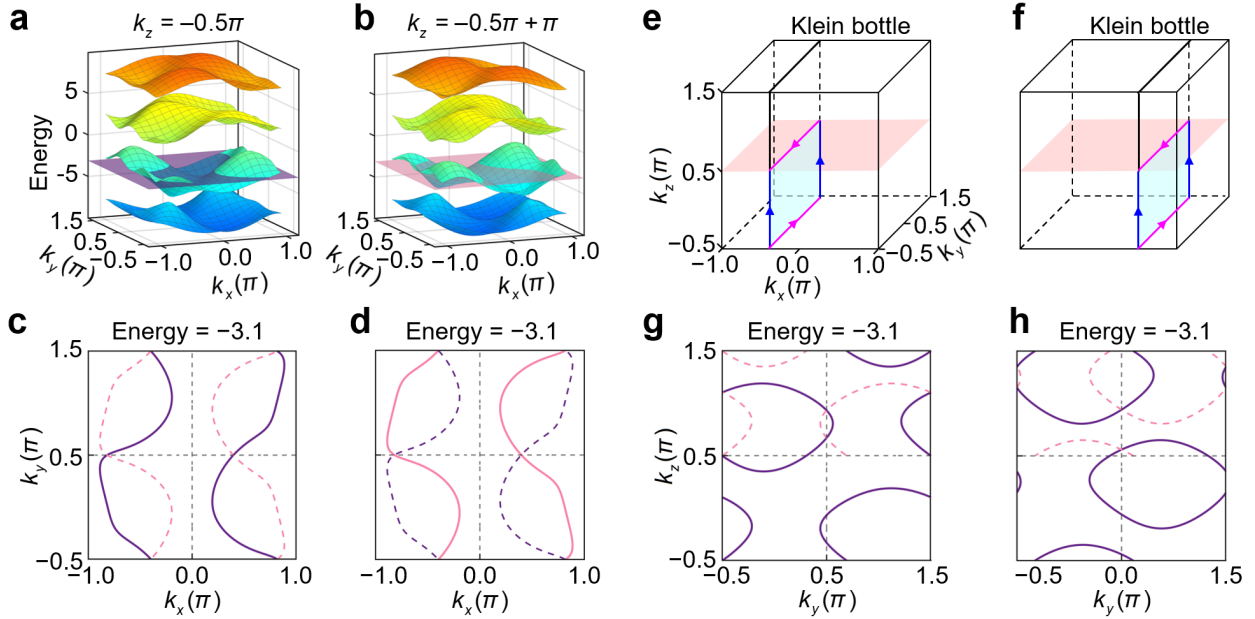

**Fig. 3 | Brillouin Klein space with momentum-space glide reflection symmetry. a, b,** Energy bands of the model with fixed $k_z = -0.5\pi$ and $k_z = -0.5\pi + \pi$. **c, d,** The constant energy cuts correspond to the purple plane in **a** and the pink plane in **b**, respectively. In **c** and **d**, the pink (purple) dashed lines can be obtained by mirroring the purple (pink) solid lines in **c** (**d**) through $k_y = 0.5\pi$ (horizontal grey dashed line), which are exactly the pink (purple) solid lines in **d** (**c**), demonstrating the momentum-space glide reflection symmetry of the band structures in the Brillouin Klein space. **e, f,** Brillouin Klein bottles (cyan planes) can be achieved by appropriately cutting the Brillouin Klein space. The marked arrow directions indicate how the boundaries with the same color will be "glued" together. **g, h,** The constant energy cuts correspond to the cut 2D BZ planes in **e, f,** respectively, which exhibit momentum-space glid reflection symmetry. For comparison, the constant energy cuts within the lower half 2D BZ is translated to the upper half 2D BZ and marked as pink dashed lines.

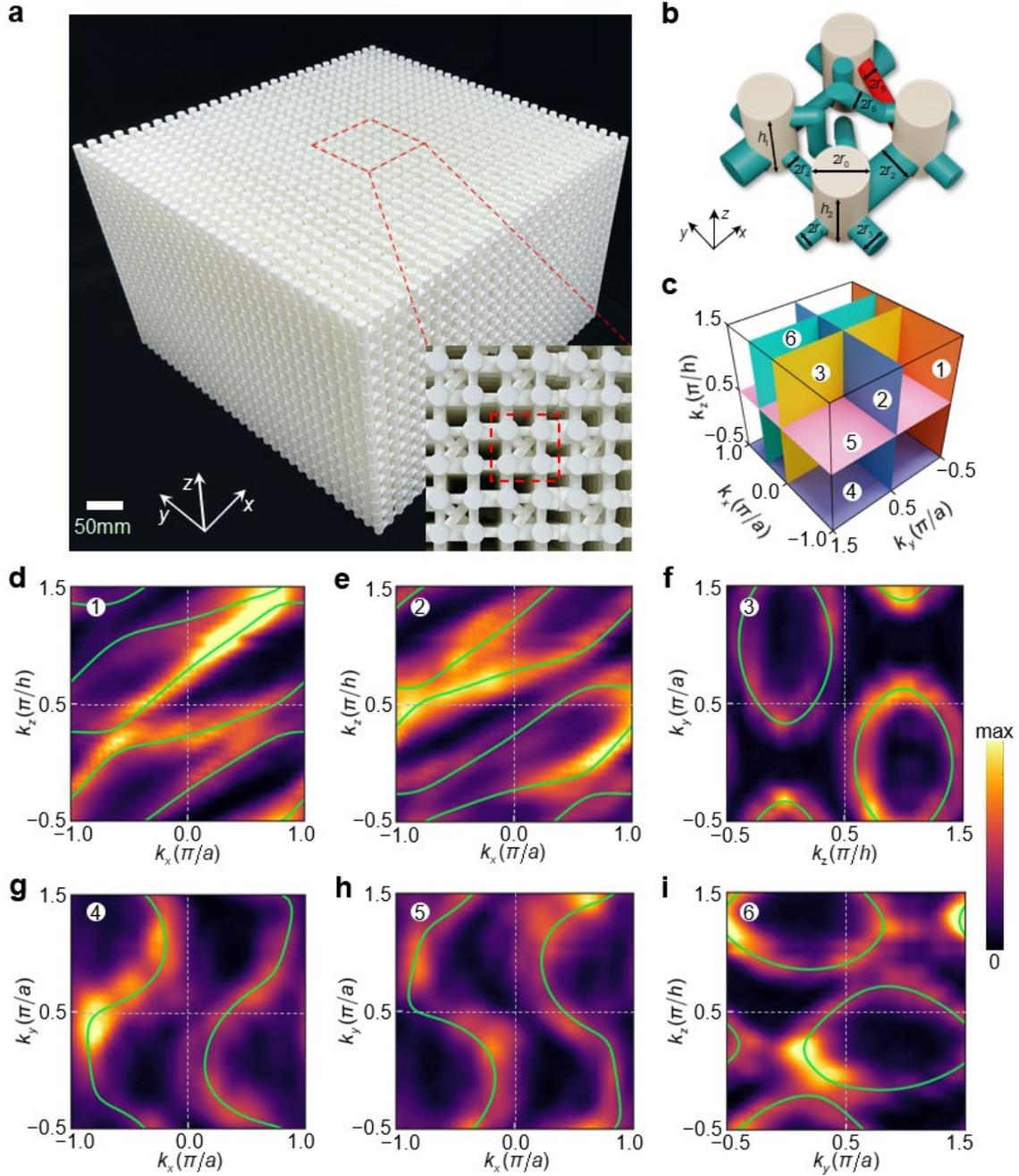

**Fig. 4 | Observation of Brillouin half-turn space and Klein space in a 3D acoustic crystal. a,** Photograph of the fabricated 3D acoustic crystal. Inset shows a top view of the sample center. The red dashed square in the inset denotes a unit cell. **b,** Unit cell of the 3D acoustic crystal consisting of four acoustic resonators (beige cylinders). The dark green (red) tubes indicate positive (negative) couplings. The lattice constants in the x-y plane and z-direction are $a$ = 48 mm and $h$ = 32 mm, and the other geometrical parameters are $r_0$ = 6 mm, $r_1$ = 2 mm, $r_2$ = 4.5 mm, $r_3$ = 3 mm, $r_4$ = 2 mm, $r_5$ = 2 mm, $r_6$ = 2.2 mm, $h_1$ = 24.5 mm, and $h_2$ = 24 mm, respectively. **c,** The six colored planes are 2D BZs which are cut from the 3D Brillouin half-turn space (①-③) or Klein space (④-⑥), respectively. **d-i,** Measured (color maps) and simulated (green lines) iso-frequency contours of the bulk band structure at 6.46 kHz that correspond to the six cut 2D BZs in **c**, respectively.

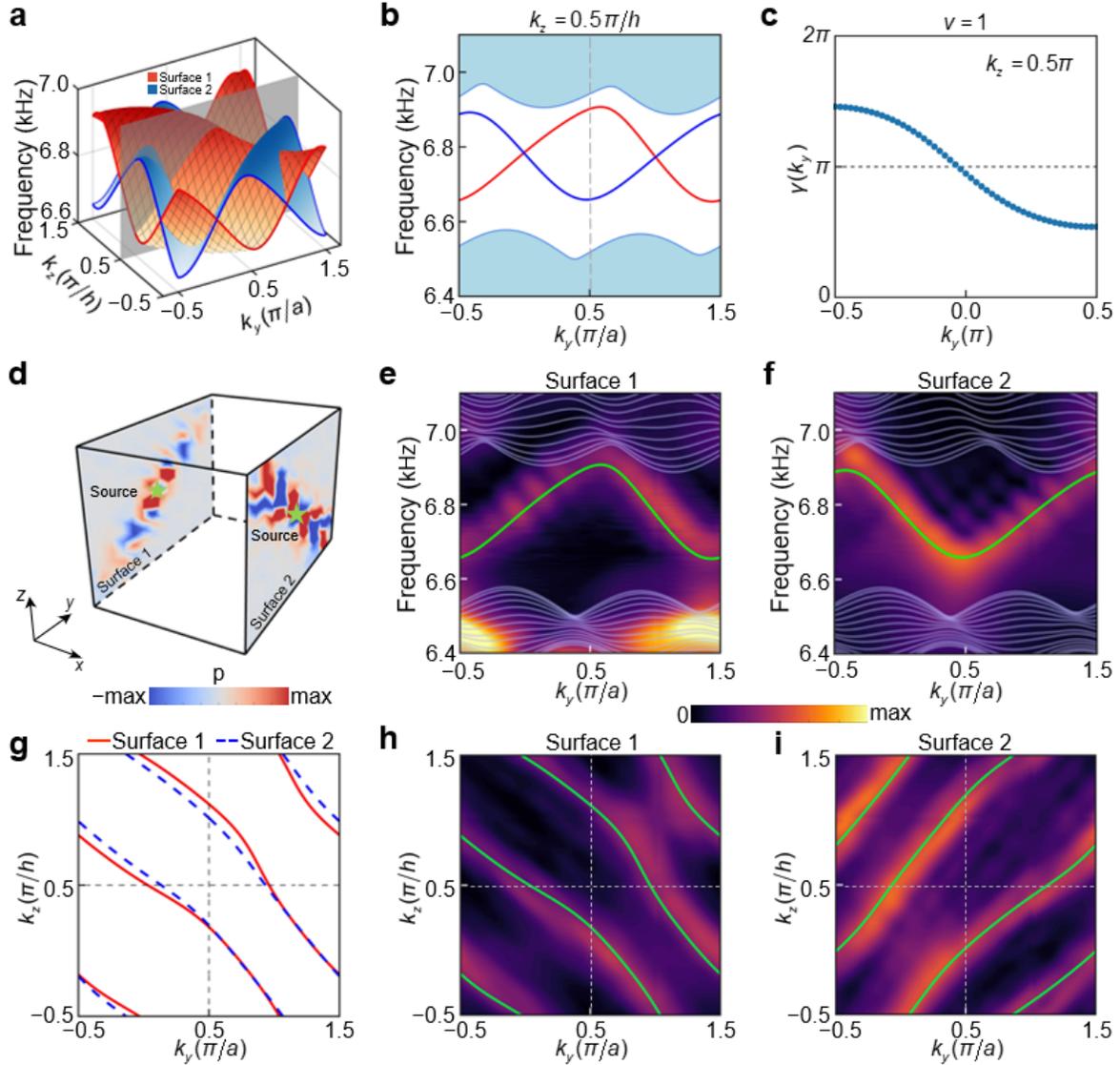

**Fig. 5 | Observation of topological surface states in a 3D Klein bottle insulator. a,** Simulated surface dispersions of surface 1 (red curved sheet) and surface 2 (blue curved sheet) that parallel to the *y-z* plane. **b,** Simulated surface dispersions with fixed $k_z = 0.5\pi/h$. The red (blue) line indicates the dispersion of surface 1 (surface 2), respectively, and the light blue regions represent the projected bulk states. **c,** The flows of $\gamma(k_y)$ for the cubic lattice model (Fig. 2a) with fixed $k_z = 0.5\pi$. **d,** Measured acoustic field distributions of the self-collimated topological surface states at 6.8 kHz on surface 1 and surface 2, respectively. Acoustic point sources (green stars) are placed at the center of surfaces 1 and surface 2 to excite the surface states. **e, f,** Measured surface state dispersions of surfaces 1 (**e**) and surface 2 (**f**) with fixed $k_z = 0.5\pi/h$. The white (green) lines indicate the simulated bulk (surface) dispersions. **g,** Simulated iso-frequency contours of surface 1 (red solid lines) and surface 2 (blue dashed lines) at 6.8 kHz which are closely related by a nonlocal twist. The reflection of the iso-frequency contour of surface 2 (blue dashed lines) through the axis of $k_z = 0.5\pi/h$ (horizontal grey dashed line) coincide with that of surface 1 (red solid lines) after a half translation $\mathcal{L}_{G_y/2}$. **h, i,** Measured (color maps) and simulated (green lines) iso-frequency contours of surfaces 1 (**h**) and surface 2 (**i**), respectively.

**Acknowledgments**

Z.G. acknowledges funding from the National Natural Science Foundation of China (grant no. 12104211, Shenzhen Science and Technology Innovation Commission (grant no. 20220815111105001), and SUSTech (grant no. Y01236148 and no. Y01236248).

**Author contributions**

Z.G. initiated the project. Z.Z., G.-G.L., L.Y. and Y.M. performed the theoretical calculation. Z.Z., X.X. and B.Y. performed the simulation. Z.Z., Z.G., G.-G.L., and Z.L. designed the experiments. Z.Z., J.W. and Z.G. fabricated samples. Z.Z., J.W., W.D., X.H. and J.L. carried out the measurements. Z.Z., L.Y., Y.M., X.X., B.Y., J.C., P.P.S., H.C., Y.Y. G.-G.L. and Z.G. analyzed the results. Z.Z. and Z.G. wrote the manuscript with input from all authors. Z.G., G.-G.L. and Z.L. supervised the project.

**Competing interests**

The authors declare no competing interests.

**Data availability**
All data that support the plots within this paper and other finding of this study are available from the corresponding authors upon reasonable request.

**End**